\documentclass[manuscript]{acmart}
\usepackage{graphicx}
\usepackage{kotex}
\usepackage{color}
\usepackage{array}
\usepackage{mathptmx}
\usepackage[utf8]{inputenc}
\usepackage{kotex}
\usepackage{subcaption}
\usepackage{enumitem}
\usepackage{comment}
\usepackage{tabularx}
\usepackage{appendix}
\usepackage{geometry}
\usepackage{multirow}

\title [Investigating Task Delegation Trend of Author-AI Co-Creation with Generative AIs]{“I’m not thinking anymore, just following the path.”: Investigating Task Delegation Trend of Author-AI Co-Creation with Generative AIs} 

\author{Yujin Kim}
\affiliation{
 \institution{Department of Computer Sciences, University of Wisconsin-Madison}
 \city{Madison}
 \state{Wisconsin}
 \country{USA}}
\email{yujinkim@cs.wisc.edu}

\author{Suhyun Kim}
\affiliation{
 \institution{Department of Artificial Intelligence Convergence, Ewha Womans University}
 \city{Seoul}
 \country{South Korea}}
\email{kimsh9977@ewhain.net}

\author{Yeojin Kim}
\affiliation{%
 \institution{Computer Science and Engineering, Ewha Womans University}
 \city{Seoul}
 \country{South Korea}}
\email{rowenna03@ewhain.net}

\author{Soyeon Lee}
\affiliation{%
 \institution{Computer Science and Engineering, Ewha Womans University}
 \city{Seoul}
 \country{South Korea}}
\email{soysilver@ewhain.net}

\author{Uran Oh}
\affiliation{%
 \institution{Computer Science and Engineering, Ewha Womans University}
 \city{Seoul}
 \country{South Korea}}
\email{uran.oh@ewha.ac.kr}

\date{Feb 2025}

\begin{document}

\begin{abstract}
This paper investigates the task delegation trends of digital comic authors to generative AIs during the creation process. We observed 16 digital comic authors using generative AIs during the drafting stage. We categorized authors' delegation levels and examined the extent of delegation, variations in AI usage, and calibration of delegation in co-creation. Our findings show that most authors delegate significant tasks to AI, with higher delegation linked to less time spent on creation and more detailed questions to AI. After co-creation, about 60\% of authors adjusted their delegation levels, mostly calibrating to less delegation due to loss of agency and AI’s unoriginal outputs. We suggest strategies for calibrating delegation to an appropriate level, redefine trust in human-AI co-creation, and propose novel measurements for trust in these contexts. Our study provides insights into how authors can effectively collaborate with generative AIs, balance delegation, and navigate AI's role in the creative process.
\end{abstract}



\maketitle

\section{Introduction}
With the advancement of AI in practical fields, studies have explored how professionals, such as educators, medical teams and annotators, delegate tasks to AI \cite{kobiella2024if, xu2023comparing, han2024teams, deng2024promoting, allen2024consent, fugener2022cognitive}.
Prior studies have addressed the causes of over/under delegation and proposed solutions for appropriate delegation. For instance, in image classification tasks, Fugener et al. \cite{fugener2022cognitive} found that human difficulty with classifying complex images can lead to excessive task delegation. This highlights the need for solutions to enhance better annotator-AI collaboration in practical scenarios.
However, in the era of creative AI, task delegation to AI in content creation remains underexplored. This gap is particularly notable as the adoption of AI by professionals for creative collaboration continues to grow.
%However, in the era of creative AI, authors' task delegation to AI in content creation tasks, where a growing number of professionals collaborate with AI, remains underexplored.
%Despite these efforts, authors' task delegation to AI in content creation tasks, where growing number of professionals collaborate with AI, remains underexplored. 
Therefore, we aim to investigate the task delegation patterns of professional authors in creation tasks. Additionally, discussions on trust have predominantly focused on tasks where correctness is well-defined, such as prediction \cite{ahn2024impact,zhang2024evaluating,ma2023should}, code generation \cite{ishaani2024evaluating,ferdowsi2024validating}, and information provision \cite{yin2024lies,leiser2024hill}. However, they remain limited for AI-driven creation tasks, where correctness is subjective or lacks clear definition. Based on the findings that trust is the most influential factor in AI delegation \cite{lubars2019ask, cvetkovic2022task}, we seek to redefine trust in creation tasks by analyzing when authors delegate the tasks to AI. Our research questions are as follows: \textit{RQ1. How authors delegate their creation tasks to generative AIs? RQ2. How does AI usage vary depending on the level of delegation? RQ3. Does the AI delegation level change in author-AI co-creation?}

With these research questions, we recruited 16 digital comic authors who work on both writing and drawing. We observed their AI usage in the drafting stage using generative AI tools such as ChatGPT 3.5 \cite{ChatGPT} and Midjourney \cite{Midjourney}. The drafting stage refers to the phase where the basic layout, composition, and flow of the comic are conceptualized and sketched. Based on a prior study \cite{ko2022we}, we divided the drafting stage into three steps; storyline development, character design, and character drawing. These steps are required drafting process over most of the content creation beyond the digital comic field. Through the study, we found that the majority of the authors considerably delegate their creation tasks to generative AIs in-the-wild. In addition, authors who delegate more to generative AIs mostly took less time for creation while asking more detailed questions to AIs. Moreover, after experiencing co-creation with generative AIs, about 60\% of authors including some authors who showed overtrust in generative AIs adjusted their delegation. Among those calibration cases, over 80\% of authors calibrated towards less delegation. The main reasons for calibration were the loss of agency, AI's unoriginal responses, and the effort to refine prompts for AI outputs which more match with their intention.

Along with these findings, we discuss how to appropriately calibrate delegation level in creation tasks based on trust calibration solutions. Moreover, we suggest how we can define trust and measure trust in creation task which do not have definitive answer. We hope our study provides groundwork about discussing how authors can collaborate with generative AIs in trustful ways while preserving their authorship and escaping from fear of losing their role in creation.
   
\section{Related Work}
\subsection{Task delegation to AI and Trust in AI}
Task delegation refers to giving a system the authority to automatically handle certain tasks. This can happen either without the user fully understanding how it works or with the user supervising the system’s actions \cite{wiener1980flight}. Due to the rapid advancement of AI, discussions around task delegation are more active than ever \cite{lubars2019ask}. For instance, Brian et al. \cite{lubars2019ask} explored how factors such as motivation, task difficulty, risk, and trust in AI, influence humans’ preferences for delegating tasks to AI. They found that trust is the most influential factor. When humans have a high degree of trust in AI, they are more willing to delegate their tasks. Trust, defined as an attitude that an agent will achieve an individual's goals in situations of uncertainty and vulnerability \cite{lee2004trust}, plays a crucial role in effective human-AI collaboration. It is essential for fostering productive human-AI interactions, which not only leverage AI assistance to human but also improve task performance \cite{bansal2021does, salimzadeh2024dealing}. As AI evolves, ensuring users place appropriate trust in the system—avoiding both overtrust and undertrust—remains a key consideration \cite{weisz2024design}. Despite these efforts, there has been limited discussion on the meaning of trusting generative AI in co-creation tasks, where definitive answers do not exist. Furthermore, how authors delegate content creation tasks to AIs remains underexplored, highlighting the need for further research in this area.

%Some of the factors that can impact on AI trust are prior experience with AI \cite{salimzadeh2024dealing}, human control over algorithmic decisions \cite{ahn2024impact}, source of training data \cite{hao2024advancing}. 
%Trust has been suggested to influence reliance; however, it alone is not enough to ensure reliance, as factors like time constraints, perceived risk, and self-confidence \cite{schoeffer2024explanations}. 
%With the rise of human-AI collaboration, escaping from users' overtrust or undertrust, and ensuring users' appropriate trust on AI gets more attention. For example, providing adaptive explanation \cite{bansal2021does} for AI output or encouraging users to review and think critically about the generative model's outputs \cite{weisz2024design} to calibrate trust have been discussed. 
%Trust is a decision by A to delegate to B some aspect of importance to A in achieving a goal \cite{wolf2024generative}. Decision delegation means entrusting others to make a decision on one's behalf \cite{candrian2022rise}.

\subsection{Human-AI Co-Creation (HACC) for Resolving Author’s Needs and Challenges}
With the development of generative AI models, AI has taken part in creative tasks, which were considered as a unique ability of humans. These days, various contents can be generated by generative AIs including text (e.g. novel \cite{yuan2022wordcraft}, advertising phrase \cite{saputra2023impact}), sound (e.g. music \cite{wang2022cps}, sound effects \cite{kang2023fall}), and image (e.g. emoticon \cite{yang2021icon}, character \cite{ruan2022anime}, human face \cite{jadhav2023high}). With the progress of AI capacity, the HACC, which means humans and AI collaborate on a shared creative product as partners \cite{rezwana2022designing}, has been increasingly studied.
For instance, researchers have explored technical gaps in authors' working practice and environment where HACC can solve \cite{inie2023designing,ko2023large,wan2023it}. %Ko et al. \cite {ko2023large} interviewed visual artists to understand how they would adopt large-scale text-to-image models to find new opportunities to support their creative works in the visual art field. 
In addition, based on these findings, HACC tools solving the gaps were suggested \cite{haoran2023magical,wang2023popblends,liu2022opal,ippolito2022creative,dang2022beyond}. Haoran et al. \cite{haoran2023magical} developed a system using AIs to help beginner artists create traditional Chinese paintings.
However, while most HACC-related studies focus on technically assisting authors, it is crucial to examine how authors in practice delegate creation tasks to generative AI to ensure more trustful HACC systems.

\section{Method}
We conducted brief interviews on authors' working practices and held hands-on AI sessions to observe their comic drafting stage with generative AIs. Based on a prior study \cite{ko2022we}, we divided the drafting stage into three steps: (1) \textbf{storyline development} that creates the overall story including choosing the genre, background, number of major characters, and their storyline; this step occurs at the beginning of comic creation, (2) \textbf{character design} that designs each character's appearance including hairstyle, clothing, facial expressions, and postures, and (3) \textbf{character drawing} that produces a representative image of the characters by applying their design to best fit the storyline.

\subsection{Participants}
We recruited 16 digital comic authors by contacting them through Instagram and snowball sampling \cite{goodman1961snowball}. We searched them by keywords such as `Webtoon’ and `Instatoon’. Specifically, we recruited digital comic authors who have used generative AI at least once and have uploaded at least one comic online. Their average age was 26.8 (\textit{SD} = 6.2) and all but two were female. The gender imbalance reflects the female-dominated digital comic industry \cite{numberOffemale}. Among 16 participants, one had more than 10 years of creating digital comics, while most had one to three years ($N$=12), and 3 participants had less than a year. Participants were compensated approximately \$37 as a reward for an in-person study. 
\begin{comment}
    The compensation was set by referring to the Korean Labor Standards Act.
\end{comment}

\subsection{Procedure}\label{procedure}
First, we conducted a brief interview to investigate how participants work during their drafting stage including consuming time and tools they use. In addition, we asked about their expectations with utilizing generative AIs during the drafting stage. Then, we held hands-on AI sessions to observe participants' generative AI usage and trends of task delegation. We asked them to conduct the three steps of comic drafting with generative AIs. Before each of the steps, we gave a tutorial to show how to use ChatGPT and Midjourney including prompting examples. %To prevent the tutorial’s effect on the main task, we requested participants to perform the task based on their usual workflow, without adhering to the tutorial.
Additionally, to facilitate their effective ideation, we provided them with paper and pens to write on freely. First, in the storyline development step, the participants were asked to draft the entire story including genre, background, number of major characters, and what happens to the characters. In this step, we set the topic of their story as ``school'' to explore the participants’ experience independently of the subject. In the character design step, the participants were asked to design the setting of a main character based on the previous step. To ensure the participants’ deep exploration for this step, we asked them to draft only one main character. In the character drawing step, the participants were asked to draw the main character they designed in the previous step. 

ChatGPT assisted with storyline development and character design steps, while Midjourney was used for character drawing step. Participants were given 10 minutes to perform the task for each of the storyline development and character design steps. But for the character drawing step, they were given 15 minutes since the inference time of the Midjourney was longer than that of the ChatGPT. We allowed participants to complete tasks earlier than the given time, or request additional time if needed. Through this, we prevented them from being bored or skipping some processes in drafting stages due to the time left or lack of time during the task. To collect usage data, we saved the usage logs including what prompts the participants gave to generative AIs, what responses they received, and how much time they took. 
After each step, we asked participants how they used generative AIs in their creation to assess their level of delegation. Furthermore, we inquired about the challenges they faced while using the AIs in their work process. Additionally, we asked participants about their future plans for utilizing generative AIs in the step to check if task delegation calibration had occurred.
At the end of the study, what role AI should play in the creation field.

\begin{figure}[hbt!]
\label{AI Task Process}
  \centering
  \begin{subfigure}{.3\linewidth}
    \centering
    \includegraphics[height = 0.95\linewidth]{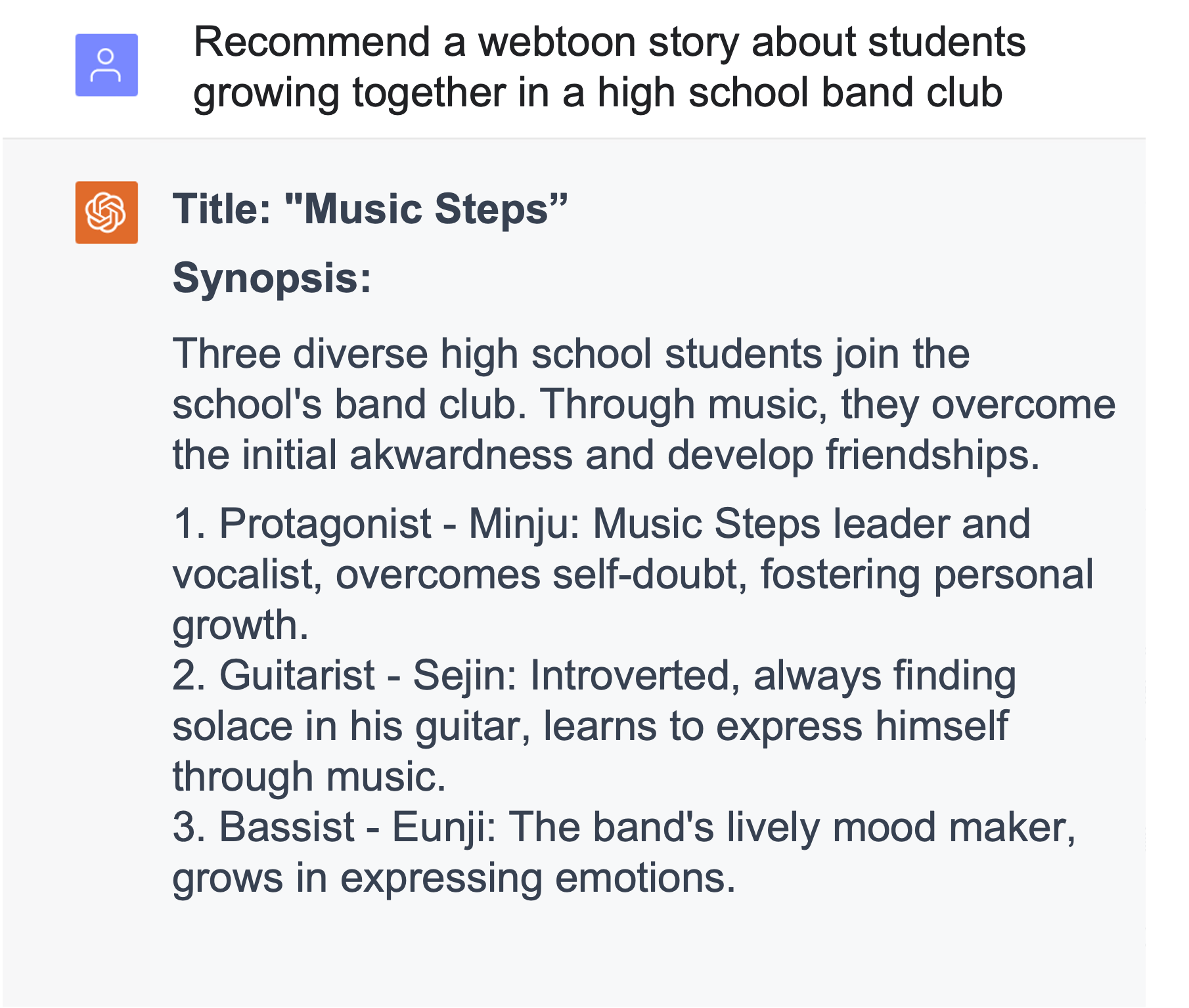}
    \caption{Storyline development}
    \label{AI Task Process - a}
  \end{subfigure}
  \hspace{1.5em}
  \begin{subfigure}{.3\linewidth}
  
    \centering
    \includegraphics[height = 0.95\linewidth]{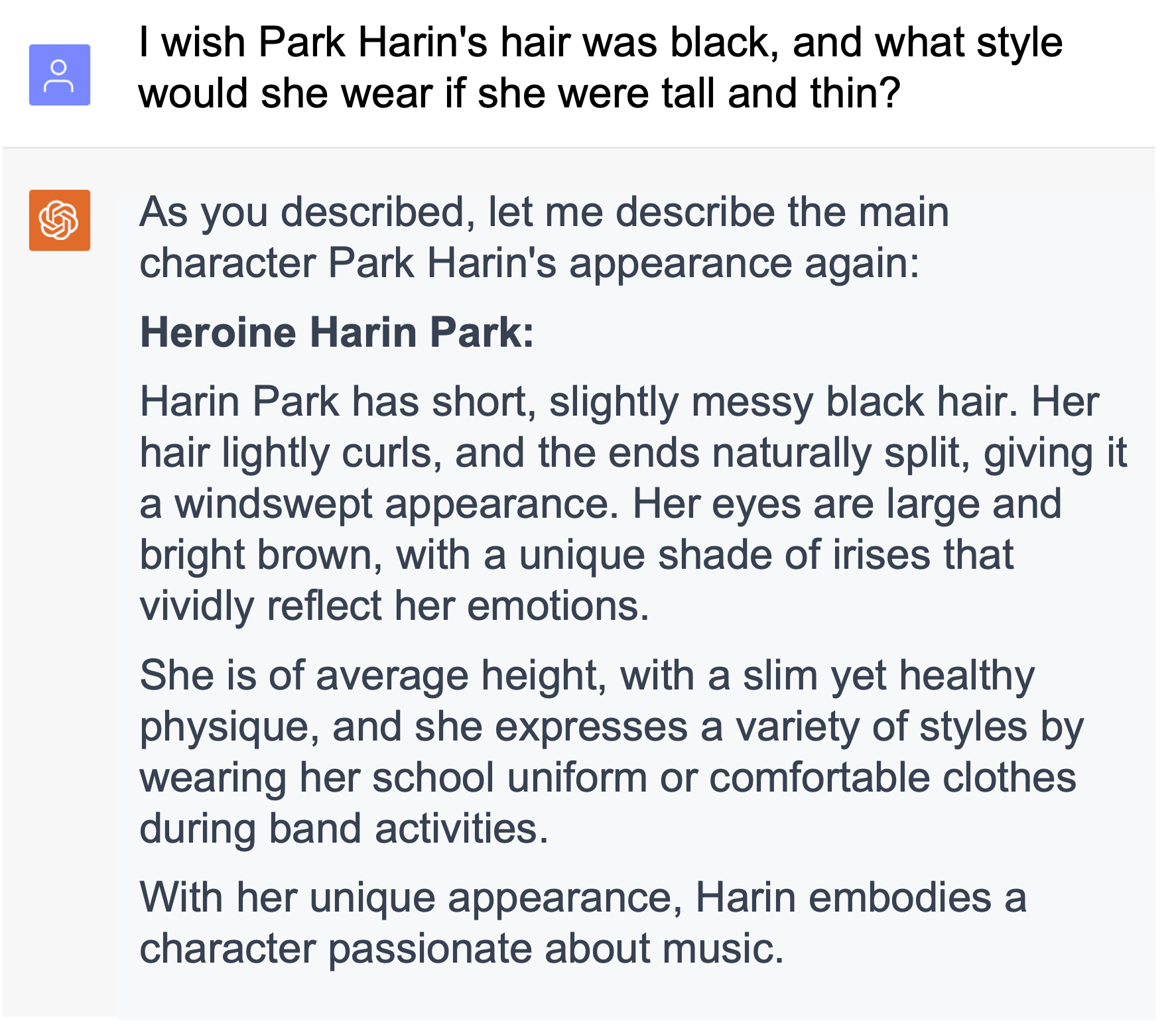}
    \caption{Character design}
    \label{AI Task Process - b}
  \end{subfigure}
  \hspace{0.0em}
  \begin{subfigure}{.3\linewidth}
  
    \centering
    \includegraphics[height = 0.95\linewidth]{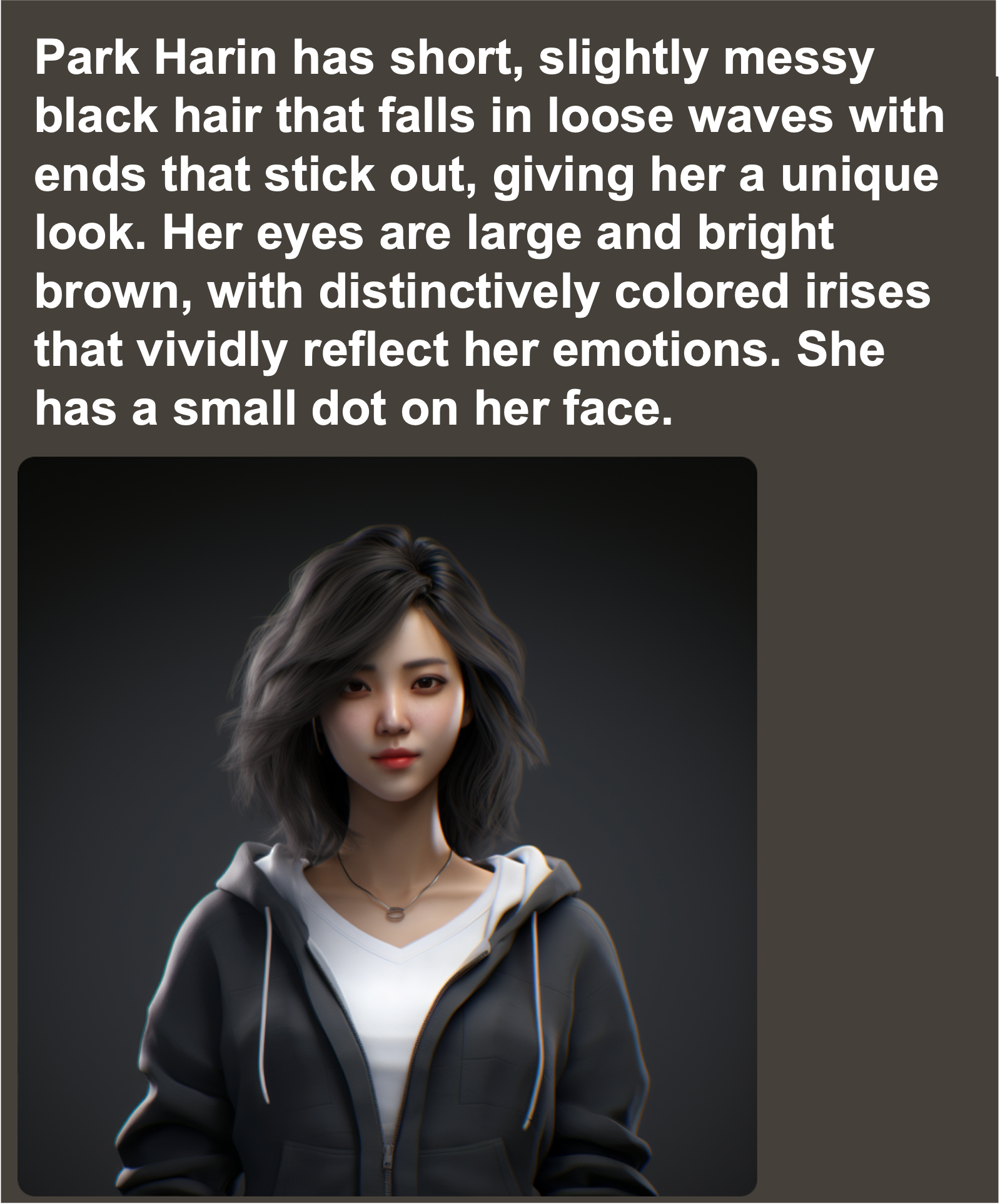}
    \caption{Character drawing}
    \label{AI Task Process - c}
  \end{subfigure}
  \caption{An example of the overall procedure of hands-on AI sessions. In (a), an author asks to recommend a webtoon storyline about students in a high school band club. In (b), she asks for recommendations for Harin’s appearance who has black hair and is tall and thin. In (c), she asks to draw Harin who has short messy black hair and a unique look with a detailed appearance.}
\end{figure}

\section{Findings}
\subsection{From Definition of Task Delegation Level to Generative AI Usage}
\subsubsection{Defining Task Delegation Level in Author-AI Co-Creation}
There have been several studies that defined the level of task delegation with an automated system such as computers \cite{endsley1999level,riley1989general,parasuraman2000model,fereidunian2007challenges}. Based on the previous research, we defined a novel task delegation level for creation task with generative AI for storyline development, character design, and character drawing steps (See \autoref{task_delegation_level}). These steps are necessitated creation process in most early stages of content creation. 

% \usepackage{multirow}
% \usepackage{graphicx}
% \usepackage{multirow}
% \usepackage{graphicx}

% \usepackage{multirow}
% \usepackage{graphicx}
% \usepackage{multirow}
% \usepackage{graphicx}
\begin{table}[hbt!]
\centering

\begin{tabular}{c|c|>{\raggedright\arraybackslash}m{0.22\textwidth}|>{\raggedright\arraybackslash}m{0.22\textwidth}|>{\raggedright\arraybackslash}m{0.22\textwidth}}
\hline
\multirow{2}{*}[-0.6em]{Entrustment}  & Full   & \multicolumn{3}{>{\raggedright\arraybackslash}m{0.66\textwidth}}{I will accept the entire \{storyline, character's traits, character's image\} \newline that AI \{writes, designs, draws\}.}      \\ \cline{2-5} 
                              & High   & \multicolumn{3}{>{\raggedright\arraybackslash}m{0.66\textwidth}}{I will slightly modify the \{storyline, character's traits, character's image\}\newline that AI \{writes, designs, draws\}.} \\ \hline
Collaboration                 & Middle & \multicolumn{3}{>{\raggedright\arraybackslash}m{0.66\textwidth}}{I will modify AI's suggestions to fit my idea.}                                                                                                                        \\ \hline
\multirow{2}{*}{Independence} & Low    & \multicolumn{3}{>{\raggedright\arraybackslash}m{0.66\textwidth}}{I will evaluate my idea using AI.}                                                                                                                                    \\ \cline{2-5} 
                              & No     & \multicolumn{3}{>{\raggedright\arraybackslash}m{0.66\textwidth}}{I will not use AI, I will do it myself.}                                                                                                                                \\ \hline
\end{tabular}%

\caption{Defined task delegation levels in Author-AI co-creation for digital comic drafting stage.}
\label{task_delegation_level}
\end{table}

\subsubsection{Generative AI Usage by Each Delegation Level}
\begin{figure}[!hbt]
\centering
\begin{subfigure}{.45\linewidth} % 너비를 조정
    \centering
    \includegraphics[width=\linewidth]{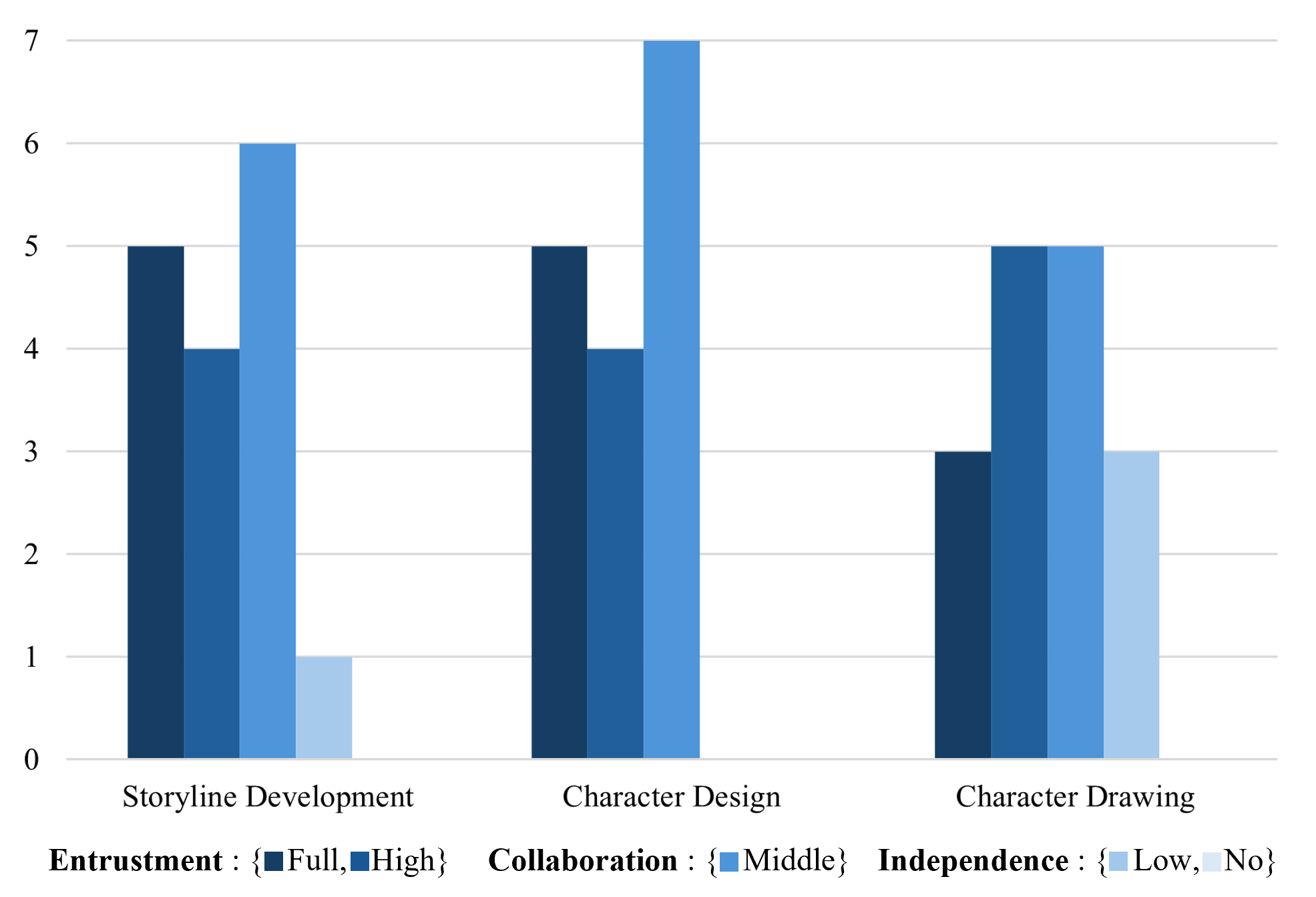} % height 대신 width를 사용
    \label{howmany}
\end{subfigure}
\hspace{1.5em} % 간격을 자동으로 조정
\begin{subfigure}{.45\linewidth} % 동일한 너비
    \centering
    \includegraphics[width=\linewidth]{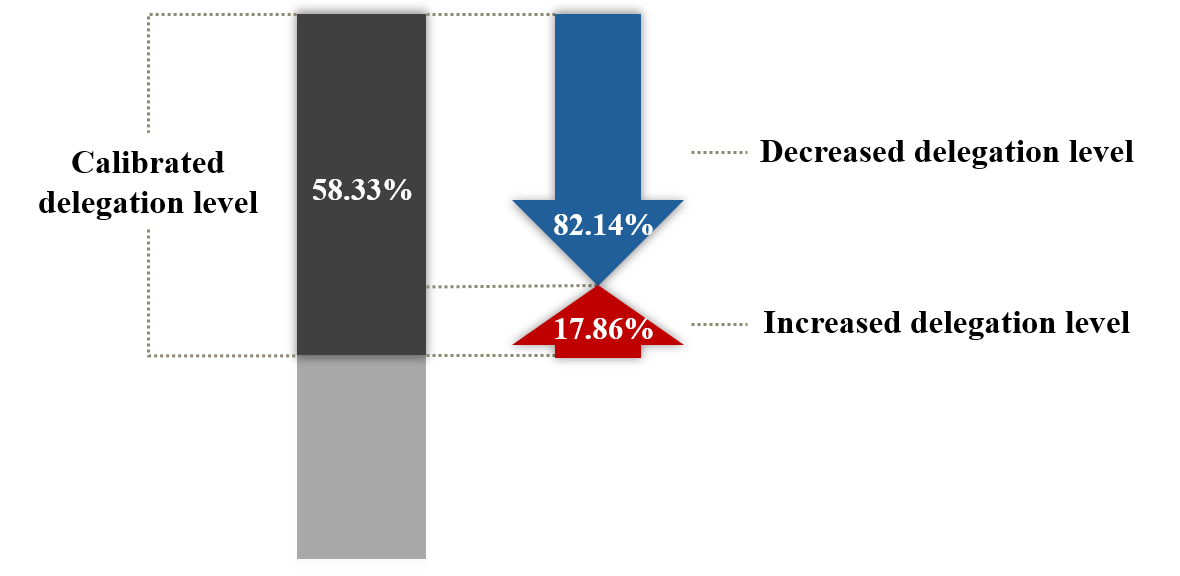}
    \label{calibration}
\end{subfigure}

\caption{The number of participants for each delegation levels (left) and the calibration trend in task delegation level (right).}
\label{Figure2}
\end{figure}

As shown in the left graph of \autoref{Figure2}, we counted the number of participants for each level of delegation. In each drafting stage, all participants delegated their creation to generative AIs and over 50\% of participants delegate the creation task with \textit{Entrustment} level (Full, High). In addition, the \textit{Collaboration} level (Middle) accounted for the largest portion across all drafting stages, while in the character drawing step, the High delegation of \textit{Entrustment} level also represented the largest portion. By each of the delegation levels, we analyzed how the AI usage differs. There were trends between \textit{Entrustment} (Full,High), \textit{Collaboration} (Middle), and \textit{Independence} (Low,No) levels.

%\textbf{When delegation level \textcolor{red}{$\Uparrow$}, time taken for creation mostly \textcolor{blue}{$\Downarrow$}}
\textbf{When delegation level increases, time taken for creation mostly decreases} 
The time taken for the creation task decreases with participants who have a higher delegation level in the storyline development step \textcolor{black}{(\textit{Entrustment}:530.22, \textit{Collaboration}:547.83, \textit{Independence}:600.0 (sec))} and character design step \textcolor{black}{(\textit{Entrustment}:405.22, \textit{Collaboration}:429.0 (sec))}. However, the trend did not hold in the character drawing step, where participants in \textit{Independence} level took the shortest time among all levels \textcolor{black}{(\textit{Entrustment}:669.38, \textit{Collaboration}:633.0, \textit{Independence}:633.0 (sec))}. 

%\textbf{When delegation level \textcolor{red}{$\Uparrow$}, length and frequency of detailed questions to generative AI \textcolor{red}{$\Uparrow$}} 
\textbf{When delegation level increases, length and quantity of detailed questions to generative AI increase} Participants who delegate more creation tasks to generative AIs tend to use longer text prompts and ask more detailed questions than those who delegate less. The average prompt length increases with higher delegation in storyline development \textcolor{black}{(\textit{Entrustment}:33.14, \textit{Collaboration}:32.35, \textit{Independence}:15)} and character design steps \textcolor{black}{(\textit{Entrustment}:29.3, \textit{Collaboration}:26.76)}.
Regarding detailed questions, the number of prompts for each type increases with higher delegation level in storyline development and character design steps. For instance, prompt types in character design step were categorized into characters' appearance (41\% of total prompts), clothing style (19\%), facial expression (9\%), height (5\%), posture (5\%), and others. For all types except clothing style, the average number of prompts increases with delegation as \autoref{table:entrustment_scores}.
In storyline development step, the types of prompt type were character settings (37\%), story narratives (36\%), ending plot (7\%), genre (3\%), background (2\%), number of characters (1\%), and others. As with the character design step, the average number of prompts for character settings, ending plots, and the number of characters increases during the storyline development step as the delegation level rises. %(See \autoref{table:entrustment_scores})

% \usepackage{multirow}
% \usepackage{graphicx}
% \usepackage{multirow}
\begin{table}[h!]
\centering
\begin{tabular}{c|c|c c c}
\hline
\textbf{Category}                                                                 & \textbf{Component}   & \textbf{Entrustment} & \textbf{Collaboration} & \textbf{Independence}  \\ 
\hline
\multirow{6}{*}{\begin{tabular}[c]{@{}c@{}}Storyline \\ Development\end{tabular}} & Character Setting    & 3.22 (1.99)                & 3.00 (2.90)                   & 3.00 (0.00)                  \\ \cline{2-5} 
                                                                                  & Story Narrative      & 2.78 (1.30)                & 3.67 (2.34)                  & 5.00 (0.00)                 \\ \cline{2-5} 
                                                                                  & Ending Plot          & 0.67 (0.50)                & 0.50 (0.67)                   & 0.00 (0.00)                   \\ \cline{2-5} 
                                                                                  & Genre                & 0.33 (0.50)                 & 0.00 (0.00)                      & 1.00 (0.00)                 \\ \cline{2-5} 
                                                                                  & Background           & 0.11 (0.33)                & 0.17 (0.41)                   & 1.00 (0.00)                 \\ \cline{2-5} 
                                                                                  & Number of Characters & 0.11 (0.33)                & 0.00 (0.11)                     & 0.00 (0.00)                    \\ \hline
\multirow{5}{*}{\begin{tabular}[c]{@{}c@{}}Character \\ Design\end{tabular}}      & Appearance           & 3.11 (2.80)                & 2.29 (1.25)                   & -                     \\ \cline{2-5} 
                                                                                  & Clothing Style       & 1.67 (2.12)                 & 1.86 (2.12)                   & -                     \\ \cline{2-5} 
                                                                                  & Facial Expression    & 0.67 (0.71)                & 0.43 (0.53)                   & -                     \\ \cline{2-5} 
                                                                                  & Height               & 0.33 (0.50)                 & 0.29 (0.49)                   & -                     \\ \cline{2-5} 
                                                                                  & Posture              & 0.56 (0.73)                 & 0.00 (0.00)                      & -                     \\ \hline
\end{tabular}

\caption{The average number of prompts for each component in storyline development and character design by delegation level. The numbers in parentheses represent standard deviations.}
\label{table:entrustment_scores}
\end{table}

%\textbf{When delegation level \textcolor{red}{$\Uparrow$}, the number of prompts to generative AIs mostly \textcolor{red}{$\Uparrow$}} 
\textbf{When delegation level increases, the total number of prompts to generative AIs mostly increases} Participants who delegate more creation tasks to generative AIs tend to use more prompts than those who delegate less. The average number of prompts increases with higher delegation levels in character design \textcolor{black}{(\textit{Entrustment}:6.56, \textit{Collaboration}:5.43)} and character drawing steps \textcolor{black}{(\textit{Entrustment}:10.88, \textit{Collaboration}:8, \textit{Independence}:6.33)}.
For the character design step, where participants used ChatGPT, we counted the number of text prompts. In the character drawing step, where Midjourney was used, we counted both the number of text prompts and button inputs (e.g., Variation\footnote{\textit{Variation} generates different versions of an image.}, Blend\footnote{\textit{Blend} merges 2–5 images into a novel image.}, etc.). The number of button inputs such as Blend, Zoom Out\footnote{\textit{Zoom Out} extends the canvas without altering the original content.}, and Pan\footnote{\textit{Pan} expands the canvas in a specific direction.} increase with higher delegation levels in character drawing \textcolor{black}{(e.g., Blend:0.33 < 0.6 < 0.63, Zoom Out:0.0 < 0.2 < 1.25, Pan:0.0 < 0.2 < 0.38)}.
However, in storyline development, the trend reverses: the number of inputs decreases with higher delegation levels \textcolor{black}{(\textit{Entrustment}:7.78, \textit{Collaboration}:8.67, \textit{Independence}:10.0)}. 

\subsection{Task Delegation Level Calibration and Expected Role of AI in Co-Creation}
\subsubsection{Task Delegation Level Calibration}
As mentioned in \autoref{procedure}, we observed the extent to which participants delegated tasks to generative AIs and, after completing the tasks, asked how much they would like to delegate in the future. Based on these data, we explored how the participants' task delegation level were calibrated. As shown in the right graph of \autoref{Figure2}, delegation levels were adjusted in 58.33\% of the 48 creative cases (16 participants’ creations across three drafting steps). Among the adjusted cases, 82.14\% showed a decrease in delegation level. The most common decreasing trends were from \textit{Entrustment} to \textit{Independence} ($N$=9), including transitions such as Full to Low ($N$=5), High to Low ($N$=2), Full to No ($N$=1), and High to No ($N$=1). Conversely, the most common increasing trends occurred within the \textit{Entrustment} group ($N$=2; High to Full).

In addition, using outcome-based trust measurement \cite{ma2024you}, we assessed participants' trust. This approach categorizes users' trust levels into appropriate and inappropriate trust, based on whether their reliance on AI responses aligns with the perceived correctness, as assessed by the author. After each step, we asked participants how they used generative AIs, whether they based their reliance on the AI’s recommendations on perceived correctness, and the challenges they encountered. Their responses were then analyzed to measure their trust levels. We found overtrust in generative AIs during in-the-wild human-AI co-creation. Among three drafting steps, the most participants ($N$=7) overtrusted generative AIs in the storyline development step. Among them, four participants decreased their delegation level. The most common decreasing trend were \textit{Entrustment} to \textit{Independence} ($N$=3;Full to Low). In the character design step, four participants overtrusted generative AIs, and two participants decreased their delegation level from \textit{Entrustment} to \textit{Independence} (Full to No) and \textit{Collaboration} to \textit{Independence} (Middle to Low) each.
In the character drawing step, one participant showed decreased delegation level, but he did not calibrate his delegation level.

\subsubsection{Causes of calibration from higher to lower delegation level}
One of the primary reasons was the losing of agency ($N$=10). Participants reported that they would not think by themselves for creation and just follow what generative AIs created in the creation process. It made them fearful of being stuck in one form, even though thinking outside the box is important for generating novel ideas. \textit{``It feels like I’m not thinking anymore, just following the path. Normally, I take time to brainstorm, but here, it goes straight to the result. It’s convenient, but also uncomfortable because I didn't use my creativity at all. I wonder, `Do I even need to do anything?’” (P13).} Furthermore, while following AI-generated outputs, some participants expressed concerns about the possibility of overlap with outputs from other users ($N$=2). %\textit{``If someone else asks ChatGPT the same thing, they'll get a similar answer. In the webtoon market, I'm worried about that could lead to overlap.” (P6)}

Another reason was that AI responses lacked originality ($N$=10), which led to disappointment as the outputs failed to meet participants' creative expectations in each drafting stage. During both the storyline development and character design steps, participants appreciated the speed and reference suggestions but were disappointed by the AI's lack of originality, perceiving the outputs as generic and not distinctive. \textit{``The setting feels basic. It seems the AI's been trained on a lot of mainstream stuff, so it doesn’t stand out as unique. The character setup didn’t feel special either.” (P11)}

The other reason is that participants had to invest significant time and effort in modifying the prompts to achieve the desired results ($N$=13). For instance, during the character drawing step, participants expected high-quality and consistent styles but were dissatisfied with outputs that deviated from their specific intentions (e.g., style or composition), leading to a cumbersome revision process. \textit{``To modify the image, I have to keep typing most of the details until I get a detailed image. I prefer drawing it myself, which feels more convenient.” (P5)} In addition, some of the participants mentioned that modifying prompts is difficult ($N$=4). 

\subsubsection{After calibration, what role should generative AI play in co-creation?} 
Based on the interview, we analyzed the future role that participants expect AI to play in the co-creation process. We categorized these roles into three: reducing workload by taking over tedious tasks, assisting with ideation by providing references, and helping to refine and specify ideas. Seven participants hoped AI would manage repetitive tasks, such as outlining, sketching, and coloring, especially during the character drawing step. Nine participants expected AI to reduce the time spent searching for references to generate novel ideas. Finally, two participants suggested AI could expand ideas based on the basic storyline and context developed by the authors. Two participants (P7, P15) mentioned all of the three roles.

\section{Discussion}
\subsection{Balancing Task Delegation in Author-AI Co-Creation} \label{discussion1}
In the wild, over half of authors delegate their creation tasks to generative AI at the \textit{Entrustment} level. Even with sufficient time, including extensions upon request, they rely on generative AIs for drafting each step, making only minor adjustments. This over-delegation has led many authors to feel a loss of agency, even questioning their role in the creation process. Such frustration mirrors what authors experience when they realize the generative AI’s abilities are comparable to their own, even to the point of replacement \cite{kawakami2024impact}. Moving forward, it is crucial to help authors establish an appropriate level of delegation to ensure effective human-AI co-existence. This balance will preserve authors' agency and ownership while fully leveraging AI technologies. Although the CHI community has explored calibrating trust in AI \cite{ma2023should, wischnewski2023measuring, zhang2024s}, the focus on the appropriate level of task delegation has been limited. By building on trust calibration studies, the community can expand these discussions.

For instance, introducing cognitive forcing functions before generative AI's creation could facilitate appropriate delegation level calibration \cite{weisz2024design, buccinca2021trust}. These processes prompt users to slow down and reconsider key decisions. One approach could involve moments where users pause to critically assess the AI’s creation, ensuring thoughtful delegation rather than over-delegation. Another solution is developing adaptive AI co-creation systems, similar to adaptive explainable AI systems \cite{bansal2021does}, that adjust humans' level of task delegation based on AI confidence. Such systems could guide users to delegate more when AI confidence is high and less when it is low, encouraging author involvement in uncertain situations. Additionally, the system could estimate users' delegation levels based on AI usage patterns, such as the average length of prompts, allowing for calibration to an appropriate level. Finally, clarifying the AI's role can help shape trust, as the trust varies depending on the defined role \cite{kim2023one}. A clear definition of the AI's role can influence task delegation by determining the level of trust and control users have over the AI at various stages of the creative process.

\subsection{Redefining Definition of Trust and Redesigning Trust Measurement}
Based on the definition of trust in AI, a person who trusts AI is someone who believes that AI will accomplish their goals overcoming uncertainty or vulnerability. For example, in a house price prediction task, if a person believes the AI model can predict the price with high accuracy, it indicates the person trust the AI. Here, uncertainty and vulnerability are the potential for inaccurate predictions. However, in a code generation task, the meaning of uncertainty and vulnerability shifts. In this case, uncertainty and vulnerability are the possibility of errors or unintended operations in the generated code. Therefore, the meaning of trust can vary depending on the specific expectations humans have of AI in each task. Our findings on the primary reasons for delegation calibration show that authors expect generative AIs to provide novel ideas and to produce responses that also closely align with their intentions in creative works. This implies that we could define trust in generative AIs for authors in creative tasks as the ability to offer original ideas and reproduce texts and images that align with the author's intention, using only minimal prompts. With further exploration within the CHI community, we could develop a more thorough definition of trust in author-AI collaboration.

If the meaning of trust can vary based on human expectations of AI, the measurement of trust should also adapt accordingly. One possible approach is to calculate the gap between human expectations and actual outcomes. For example, we can measure the gap between the expected originality of a response and the actual originality of the outcome. Similarly, the gap could reflect the expected closeness of the AI’s response to the author’s intention compared to the actual difference between them. A large gap might indicate either overtrust or undertrust, while a smaller gap suggests an appropriate level of trust. Since trust can differ depending on the task and the user's expectations, using different measurements of trust may be more suitable for author-AI co-creation tasks. These measurements could complement outcome-based and behavior-based metrics, which are common methods for assessing trust \cite{ma2024you}, offering a more comprehensive analysis of trust in author-AI collaboration.

\section{Limitation \& Future Work}
%% 기존에 쓴 내용
In this study, to explore authors' task delegation to generative AI, we employed general-purpose AI tools that were not specifically designed for digital comic authors. This was due to the limited availability of dedicated tools for the comic drafting stage. Consequently, this experimental setup made it challenging to fully integrate the generative AIs into the drafting process in real-world settings. Therefore, as future work, we could develop and adopt a HACC tool to investigate their practices in using generative AI tools in more realistic situations.

\section{Conclusion}
In this paper, we examined the trend of digital comic authors' task delegation to generative AI in author-AI co-creation tasks. We defined task delegation levels and investigated the differences in generative AI usage at each level. Additionally, we analyzed the calibration trends in task delegation and explored the factors influencing such calibration. Based on our findings, we proposed solutions for appropriately calibrating task delegation levels and redefined the concept of trust in generative AI’s creation tasks. Furthermore, we introduced a novel design for measuring trust in author-AI co-creation tasks based on this redefined concept. In summary, this paper offers a timely and unique perspective on task delegation, delegation calibration, and trust in co-creation with generative AIs. We hope our research will contribute to the CHI community’s understanding of human-AI collaboration and inform future studies on this evolving field.

\bibliography{ref}
\bibliographystyle{unsrt}
\end{document}